# Optical computing by injection-locked lasers

T. von Lerber[1,2]*, M. Lassas[2], Q. T. Le[1], V. S. Lyubopytov[1,3], A. Chipouline[1], K. Hofmann[4], F. Küppers[1,5]

1. Photonics Lab, Technische Universität Darmstadt, Germany
2. Department of Mathematics and Statistics, University of Helsinki, Finland
3. Telecommunication Systems Dept., Ufa State Aviation Technical University, Russian Federation
4. Integrated Electronic Systems Lab, Technische Universität Darmstadt, Germany
5. College of Optical Sciences, University of Arizona, USA
* Corresponding author

**A programmable optical computer has remained an elusive concept. To construct a practical computing primitive equivalent to an electronic Boolean logic, one should find a nonlinear phenomenon that overcomes weaknesses present in many optical processing schemes. Ideally, the nonlinearity should provide a functionally complete set of logic operations, enable ultrafast all-optical programmability, and allow cascaded operations without a change in the operating wavelength or in the signal encoding format. Here we demonstrate a programmable logic gate using an injection-locked Vertical-Cavity Surface-Emitting Laser (VCSEL). The gate program is switched between the AND and the OR operations at the rate of 1 GHz with Bit Error Ratio (BER) of $10^{-6}$ without changes in the wavelength or in the signal encoding format. The scheme is based on nonlinearity of normalization operations, which can be used to construct any continuous complex function or operation, Boolean or otherwise.**

The history of computing has witnessed many techniques to perform arithmetic, where the development has progressed through mechanical devices to electronic circuits, and to quantum computers[1–3]. In the past, the computers were predominantly dedicated devices, many of them analogue calculators with fixed programming, which efficiently solved a given problem. Later, calculators progressed towards general-purpose digital computers that could be programmed and re-programmed over again. On the leading edge of computing, the competition between the dedicated one-solution devices versus the general-purpose computers still exists.

Our modern life is enriched by electronic computing devices, such as personal computers, which each may contain several billion transistors. In contrast, we rarely, if ever, meet an optical computer that would perform even a few Boolean operations. The challenge is the nature of photons that do not interact with each other in the vacuum but require a nonlinear medium. Optical and electro-optical computing has been attempted for some decades[4,5], yet, without a practical implementation of a one. The difficulty has been to economically cascade operations, which require a logic-level restoration, an unaltered operating wavelength, and an unmodified encoding format when passing the signal from a component to another[6]. The search for and discussion of an optical transistor, a figurative optical building block that would imitate the electrical transistor in its versatility, has remained active[7,8].

Optical transistors have been studied in a number of arrangements, such as in atomic ensembles[9–11] and in a semiconductor microcavity[12]. Optical gates and switches have been proposed by use of various nonlinear components, such as self-electro-optic effect devices[13,14], semiconductor optical amplifiers[15–19], and waveguides and resonators[20–24]. Also, metamaterials have been proposed for both analogue and digital computations[25–27].

Optical computing is predominantly used for linear operations[28,29], often accompanied with spatial light modulators[30]. Recently, optical computing has been envisioned, not as a replacement of electronic ones, but as a complementary technology in fields of optical networking[31], supercomputing[32], reservoir computing[33,34], and quantum computing[1].

In this Article, we show, for the first time, an all-optical computing primitive based on normalization operations of injection-locked lasers. We explain how normalizations can be used to construct an arbitrary continuous function $f: \mathbb{C}^n \to \mathbb{C}^n$, with any given accuracy (Supplementary Equations 1 and 2) and we demonstrate a VCSEL based all-optical programmable AND/OR logic, whose mode of operation is changed at a rate of 1 GHz with BER of $10^{-6}$ without changes in the operating wavelength or in the signal encoding format, as required for practical cascadability. We study cascadability also with an all-optical Conway's Game of Life (GoL) and simulate 100 generations of error-free gaming. The scheme can be tailored for different signal encoding formats; the Boolean logic demonstration and the GoL simulation are performed with phase and amplitude encoded signals, respectively. As a practical note, semiconductor lasers are practical nonlinear elements due to a small footprint and scalable fabrication. Finally, the scheme can be generalized beyond lasers and optics, because the phase synchronization is a fundamental property of all self-sustained oscillators.

**Operating principle**

Oscillators and oscillatory systems emerge in various shapes and forms, ranging from tidal waves to electrical circuits and down to atomic particles. Self-sustenance of an oscillator means that it has a nonlinear damping, which returns a system back to a stable oscillation regime after a perturbation, as described by the van der Pol equation. Common to all self-sustained oscillators is that they are non-conservative and consume energy to maintain a stable limit cycle. For instance, a push on a base of a mechanical metronome affects the swing of the pendulum bob. If the perturbation does not repeat the metronome returns back to its original ticking frequency, yet, the stabilization consumes energy that will be taken from the spring.

Self-sustained oscillators may synchronize with an external force, given that the frequency of the perturbation is within the locking range of the oscillator (inside the Arnold's tongue). Especially, a weakly coupled slave oscillator is known to adopt the phase of the master while maintaining its original amplitude, a phenomenon that is regularly witnessed, for instance, in synchronized pendulum clocks or in injection-locked lasers. Under certain conditions the phase synchronization can be mathematically expressed as a normalization operation $y = p\, x\, /||x||$, where $x$ and $y$ are complex amplitudes of the master and slave oscillations, respectively; and $p$ is a dimensionless amplification factor determined by the slave. Figure 1a schematically illustrates amplitudes of a master (blue line) and a weakly-coupled slave (red line). After a perturbation, the amplitude of the slave returns to a constant while the phase remains locked with the master. The normalization operation can be further rendered into the signum function (see Fig. 1b) if the phase $\Phi$ of the master equals to an integer of $\pi$.

In this Article we primarily discuss lasers, yet, the same principles could be applied to other types of self-sustained oscillators[35]. Our treatment focuses on optical systems with a single linear state of polarization, yet we readily acknowledge the possibility for two orthogonal states of polarizations in which case the signals would be expressed with two-dimensional complex amplitude vectors and normalizations of the same (see Supplementary Equations 1).

In a suitable topology, a collection of normalization operations can be used to approximate an arbitrary continuous function $f: \mathbb{C}^n \to \mathbb{C}^n$, as will be rigorously derived in the Supplementary Equations 2. In practice, this means that one may construct any conceivable logic operation or arithmetic, binary or otherwise. As a simplified example of functions depending on one variable, we can use the fact that any continuous real-valued function $f: \mathbb{R} \to \mathbb{R}$, can be approximated by a function $\phi: \mathbb{R} \to \mathbb{R}$, that is a sum of signum-functions. More precisely, for arbitrarily large $L > 0$ and arbitrarily small $\varepsilon > 0$ there exists a function $\phi$ of the form $\phi(x) = \sum_{k=-N}^{N} b_k \operatorname{sgn}(x - ka) + c$ such that for all $x$ satisfying $|x| < L$ we have $|f(x) - \phi(x)| < \varepsilon$. Here, $x$ is a real-valued input amplitude of the master (assumed master phase $\Phi = 0$ or $\pi$), $a$ determines the granularity of the approximation, $2N+1$ is the number of signum operators, $b_k$ are the weights, and $c$ is an offset (see Fig. 1c). In the physical world, this can be translated into a case, where a number of self-sustained oscillators are synchronized by a master and the input signals are combined with respective biases (see Fig. 1d). The slave oscillator outputs are weighted and summed together. The weighing can be realized, e.g., with varying coupling strengths or signal attenuations. It is assumed that the master and the slave lasers have the same angular frequencies, the signal phases stay controlled, and the oscillators in a cascaded system have a nonreciprocal flow of information, i.e., the master controls the slaves but not the opposite. Unlike most other optical signal processing schemes, we make no assumptions concerning the signal modulation format or the form of logic.

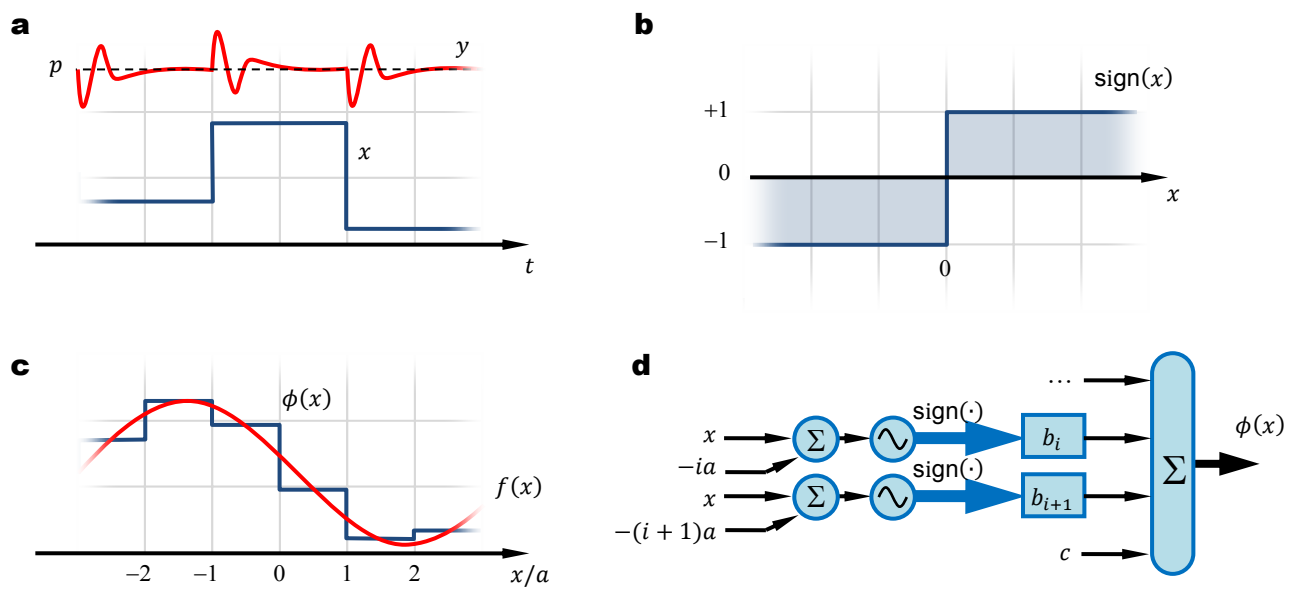

**Figure 1 | Principle of operation. a,** Slave oscillator amplitude |*y*| (red line) approaches a constant *p* at steady-state regardless of fluctuations in the master |*x*| (blue). **b,** Normalization operation is akin to signum-function for a real-valued input. **c,** A collection of signum-functions can be used to approximate an arbitrary continuous real function. **d,** a schematic illustration of a physical realization using multiple oscillators.

## All-optical programmable logic

A programmable AND/OR gate is an example of a Boolean logic that consists of an injection-locked laser with three optical inputs and an optical output. Inputs A and B and a bias X are combined at the laser and are assumed to carry equal optical power. The signals are phase encoded such that the binary numbers of 1 and 0 relate to phase-shifts of 0 and $\pi$ and complex electric field amplitudes of +1 and −1, respectively. When the bias, i.e., the program bit is set to 0 (hence the phase shift is π and the related electric field amplitude is −1), the normalization operation will produce the Boolean **AND** operation, while the bias bit of 1 will produce the **OR** operation as shown in Table 1.

**Table 1 | Truth table of a programmable AND/OR gate.** The table contains symbol values, accompanied complex electric field amplitudes, the sum of the electric fields, and the normalized output; the normalization operation is denoted as $(x)_p^0 \equiv px/\|x\|$.

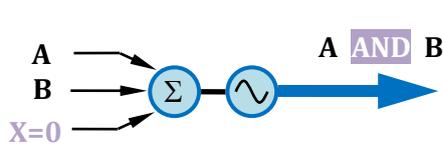

| A | B | X | $E_A$ | $E_B$ | $E_X$ | Σ | $(\Sigma)_1^0$ | AND |
|---|---|---|---|---|---|---|---|---|
| 0 | 0 | 0 | −1 | −1 | −1 | −3 | −1 | 0 |
| 0 | 1 | 0 | −1 | +1 | −1 | −1 | −1 | 0 |
| 1 | 0 | 0 | +1 | −1 | −1 | −1 | −1 | 0 |
| 1 | 1 | 0 | +1 | +1 | −1 | +1 | +1 | 1 |

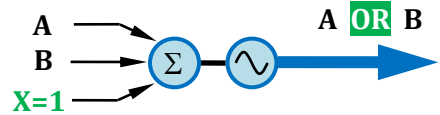

| A | B | X | $E_A$ | $E_B$ | $E_X$ | Σ | $(\Sigma)_1^0$ | OR |
|---|---|---|---|---|---|---|---|---|
| 0 | 0 | 1 | −1 | −1 | +1 | −1 | −1 | 0 |
| 0 | 1 | 1 | −1 | +1 | +1 | +1 | +1 | 1 |
| 1 | 0 | 1 | +1 | −1 | +1 | +1 | +1 | 1 |
| 1 | 1 | 1 | +1 | +1 | +1 | +3 | +1 | 1 |

We emulated the combined three signal sum input with a tandem of an external cavity laser (ECL), an electro-absorption modulator (EAM), and a phase modulator (PM) that together produced the desired multilevel hybrid encoding equivalent to the superposition of A, B, and X (see Table 2 and Extended Data Fig. 1). The bit streams of A and B were each given 128 randomly generated bits (not using linear feedback shift registers, but a computer generated pseudo-random logic gate sequence) and the bias X was altered between 0 and 1 for each operation. The respective multilevel phase-modulated signal had relative power levels of 1 and 9, i.e., the extinction ratio (ER) of 9.5 dB, and the phase shifts of 0 and π. The EAM and PM were each controlled by separate bit pattern generators (BPG) that provided independent bit patterns at a rate of 1 Gbit/s. The BPGs were synchronized in time such that the modulated optical amplitude and phase symbols overlapped. The PM output was coupled to an erbium-doped fibre amplifier (EDFA), an optical filter (OF), and a variable attenuator (ATT) in order to control and optimize the optical power of the seeding signal.

**Table 2 | Symbols and electric fields of emulated the three signal sum input.**

| bit # | A | B | X | $E_A$ | $E_B$ | $E_X$ | sum field | power | phase |
|---|---|---|---|---|---|---|---|---|---|
| 1 | 1 | 1 | 0 | +1 | +1 | −1 | +1 | 1 | 0 |
| 2 | 1 | 0 | 1 | +1 | −1 | +1 | +1 | 1 | 0 |
| 3 | 0 | 1 | 0 | −1 | +1 | −1 | −1 | 1 | π |
| 4 | 1 | 1 | 1 | +1 | +1 | +1 | +3 | 9 | 0 |
| 5 | 1 | 0 | 0 | +1 | −1 | −1 | −1 | 1 | π |
| 6 | 0 | 0 | 1 | −1 | −1 | +1 | −1 | 1 | π |
| … | … | … | … | … | … | … | … | … | … |
| 128 | 1 | 1 | 1 | +1 | +1 | +1 | +3 | 9 | 0 |

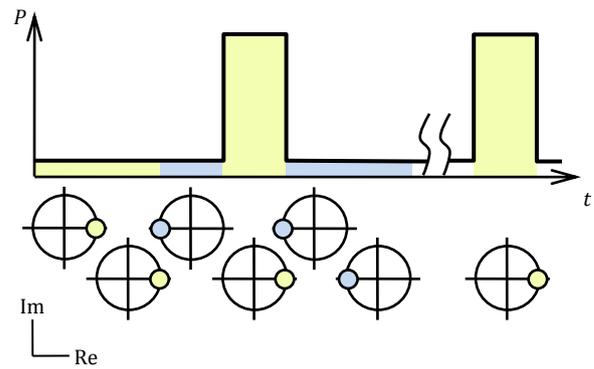

The seeding signal was coupled into the slave VCSEL via an optical circulator (OC), which directed the returning signal into a second EDFA and a second OF. The phase of an optical signal cannot be measured directly from the optical power, but it has to interfere with some known reference, which in our case was a preceding symbol. Thus, we measured a differential phase of pulses in the spirit of differential phase-shift-keyed (DPSK) demodulated output using a delay-line interferometer (DLI) with free spectral range of 1 GHz. A single output of the DLI was measured with a p-i-n photodiode that was connected to a bit error rate tester (BERT). The expected bit sequence of the DPSK modulated output was pre-programmed into the BERT, which counted the errors.

In this experiment, the slave VCSEL was pumped slightly above its lasing threshold to enable maximal speed of operation. As known from the literature of injection-locked semiconductor lasers[36] the optimal regime of injection locking, and thus normalization operations, was obtained when the detuning between ECL master and VCSEL slave approached zero.

We optimized the injection locking by setting the wavelength of ECL close to the free-running wavelength of VCSEL and then fine-tuned the lock by adjusting the pump current. Depending on both the pump current and the resulting detuning, an optimal optical injection power was set to minimize the BER of the received demodulated DPSK signal.

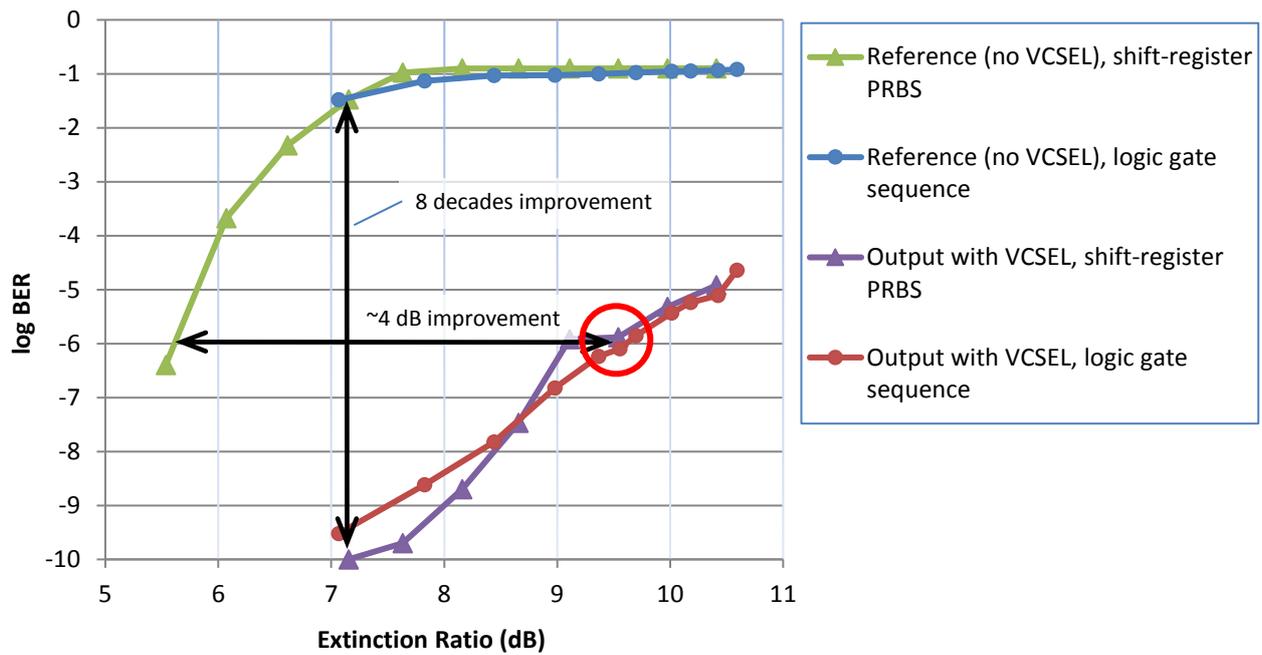

**Figure 2 | Measured BER with and without slave VCSEL for varying ER.** Red circle: logic gate BER.

The slave VCSEL's ability to quench the amplitude modulation was measured first with a shift-register-based PRBS where the EAM and the PM were fed with dissimilar word lengths of $2^9 - 1$ and $2^7 - 1$, respectively. The ER was gradually increased and the BER was measured with and without the presence of the slave VCSEL (see Fig. 2). In absence of the VCSEL the BER $> 10^{-3}$ when ER was at 6.5 dB, and when ER was above 7.5 dB, the BER became saturated. As was evident, the uncompensated DPSK receiver performance was severely compromised with an increase of the amplitude modulation. When the slave VCSEL was present, the BER was improved by about 8 decades at ER = 7.0 dB; or at fixed BER of $10^{-6}$ the improvement in ER sustainability was about 4 dB. All BER measurements were performed under constant receiving power. These results are in agreement with experiments made with an edge-emitting semiconductor laser[37]. The BER improvement is a proof of a successful injection-locking of the VCSEL. We obtained similar BER readings when the shift-register PRBS was changed to logic gate sequence feed of Table 2. As noted above, the emulated combination of inputs A, B, and bias X had the ER of 9.5 dB, which in the current setup translated in a BER of $10^{-6}$. Should the AND/OR gate be combined with a NOT operation, which in the present case is a phase shift of $\pi$, a complete functional set of Boolean operations is obtained. The AND/OR gate is analysed also mathematically in Supplementary Equations 3.

**Cascaded operations in cellular automaton**

A general-purpose computing device, optical or not, should perform numerous operations in cascade without accumulation of errors. For various reasons, this requirement has been perceived as the Achilles' heel of optical information processing, chiefly due to challenges in logic-level restoration that debilitate cascadability[38]. Also, many current optical processing schemes are based on nonlinear effects that shift the signal wavelength or change the signal encoding format, which complicates practical circuit design.

We study cascadability with a GoL, a Turing's Universal[39] cellular automaton that is played on a two-dimensional grid of cells. The Game additionally provides a demonstration of a non-Boolean logic and use of amplitude encoded signals. Rules of the Game, the optical design of a cell, and the used simulation model, including the rate equations are described in Methods section.

The initial pattern is chosen such that it is known to pulsate in three generations (see Extended Data Fig. 4) and when properly played, the living cells are confined within a limited area. For each generation, the complex electric field amplitudes of the previous generation are read and new states of the cells are computed. Importantly, the complex electric fields are not regenerated at any stage, which allows accumulation of phase and amplitude errors. An important feature of the setup is that the same bias is shared between all cells. A common bias is a key for the coherence and cascaded operations. We simulated evolution of the Game. During the course of 100 generations, we observed no degradation in gaming performance.

## Discussion

We have developed an all-optical computing scheme based on normalization operations of injection-locked lasers and provided a rigorous mathematical proof that the principle can be used to approximate any continuous one or multivariable complex-valued function. The scheme offers various possible realizations of the universal computing device and we show universality by two examples, one for phase and another for amplitude encoded signals with the programmable AND/OR gate and the GoL cellular automaton, respectively. We believe this Article offers the first demonstration of all-optical programmable logic where the control signal shares the same wavelength, the same optical power, and the same signal encoding format than the input signals. Thus, the output of a stage can be used for on-the-fly programming of the next. From a manufacturing perspective, the semiconductor lasers, specifically the VCSEL used here, are an attractive choice because they offer scalable wafer level fabrication and integration. A direct-modulated VCSEL-based optical link has been demonstrated elsewhere[40] to operate at a data rate of 71 Gb/s, which suggests that all-optical information processing could exceed the speed of CMOS electronics.

In literature, optical logic has been discussed to bear a number of so-far unsolved obstacles[6]. Here, we have shown that cascadability and logic level restoration of signals are feasible. Also, the utilization of phase modulation instead of amplitude modulation improves the resilience to optical losses, as known in the art. Difficulties with coherent processing schemes known from other applications are less of an obstacle, because we envision this scheme will primarily be used with photonic integrated circuits.

**Acknowledgements**

The research of ML was partially supported by the Finnish Centre of Excellence in Inverse Problems Research and Academy of Finland (projects 284715 and 303754). VSL acknowledges the support of the DAAD and the Ministry of Education and Science of Russian Federation ("Michail Lomonosov" programme).

**Author contributions**

TvL conceived the principle together with ML, FK, and KH. Programmable gate and GoL were co-developed by ML and TvL, the former formulated them mathematically and the latter translated them into optical designs. TvL and ML derived the steady-state laser solution and the universal approximation, respectively. QTL and VSL built the measurement setup. The programmable gate was measured by VSL who analyzed the results together with AC and FK. QTL programmed the GoL simulation algorithm. The manuscript was written by TvL with input from all authors.

**Additional information**

Supplementary information is available. Correspondence and requests for materials should be addressed to TvL (tuomo.lerber@gmail.com), ML (matti.lassas@helsinki.fi), or FK (franko.kueppers@optics.arizona.edu).

**Competing financial interests**

The authors declare no competing financial interests.

**Methods**

***Rules of Conway's Game of Life.*** The game progresses in discrete time steps such that the states of cells (living or dead) determine the outcome of the next generation. A GoL cell obeys three simple rules, which determine its fate on the next generation: If the cell is alive, then it will stay alive if it has either 2 or 3 living neighbours. If the cell is dead, then it will revive only if it has exact 3 living neighbours. If the cell doesn't stay alive or revive, it will be dead.

Mathematically, these rules can be translated into a summation of two signum functions, both being dependent on the number of living neighbours. The first signum function changes from negative to positive when the number of living neighbours is 1.5 (2.5) for a living (dead) cell. The second signum function changes from positive to negative when the number of living neighbours is 3.5. The optical power of a simulated cell output is depicted in Fig. 3a for increasing number of living neighbours. The increments occur in 2 ns intervals and the change in the number of living neighbours is assumed instantaneous. As evident, at steady-state the cell obeys the rules of the game.

***Optical design and simulation of Conway's Game of Life.*** A schematic optical implementation of the GoL cell is depicted in Extended Data Fig. 2. Component-wise the cell consists of a pair of lasers, four couplers, phase control, and a number of attenuators. At steady-state, each laser emits a constant optical power of 1 and the total output of the cell is thus 2 or 0 for states of alive and dead, respectively. For sake of simplicity and without the loss of generality the optical media and the couplings are assumed lossless.

The cell output power is divided evenly between nine paths, one for feedback and eight for feeding the neighbours – hence a 2x9 coupler – resulting in an optical power of $X_1 = x_1 x_1^* = 0.22$ for each instance when the cell is alive, where $x_1$ is the electric field amplitude of the living cell. At input, an 8x2-coupler combines optical fields of $n = \{0..8\}$ living neighbours and provides a signal $s = nx_1/\sqrt{16}$ for both arms. The electric field amplitude of the feedback $e = \{0, x_1\}$. The input side has also biasing signals, driven by respective values of $-2.5x_1$ and $-3.5x_1$. It is assumed that all bias signals across the cells have a common narrow bandwidth light source and the phases stay controlled. The bias and the feedback paths have constant attenuations of $-12$ dB in order to equalize the scales of power levels with the signal *s*.

The 3x1 coupler at the upper optical path of Extended Data Fig. 2 (marked with orange) combines the signals of the living neighbours *s*, the attenuated feedback *e*, and the attenuated bias $-2.5x_1$. If the cell was dead ($e = 0$) in the previous iteration, then the sign of the output electric field of the 3x1 coupler is negative when $n \leq 2$, and positive otherwise. If the cell was alive ($e = x_1$) then the sign of the field amplitude is negative when $n \leq 1$, and positive otherwise. The light is injected into a laser that normalizes the signal. Its steady-state electric field amplitude can be expressed as

$$E_a = \begin{cases} -1 & \text{if } (n \leq 2 \wedge e = 0) \vee (n \leq 1 \wedge e = x_1) \\ +1 & \text{otherwise} \end{cases}$$

The 2x1 coupler at the lower optical path of Extended Data Fig. 2 (marked with green) combines the signals of the neighbourhood *s* and the attenuated bias $-3.5x_1$. The output phase of the laser is shifted by $\pi$ phase shifter, which inverts the sign of the steady-state electric field

$$E_b = \begin{cases} +1 & \text{if } n \leq 3 \\ -1 & \text{otherwise.} \end{cases}$$

The normalized electric fields are combined at the output by the 2x9 coupler.

The lasers of the GoL cells are modelled using differential rate equations of solid-state lasers[41,42] with following laser parameters: Linewidth enhancement factor $\alpha = 0$ (i.e., the laser was treated as a solid-state laser, such as a Nd:YAG, not as a semiconductor one due to large dynamic range of 22 dB at the input of the upper path laser), injection current $\mu = 2.0$, decay rate of the cavity electric field $\kappa = 10^{12}$ s$^{-1}$, decay rate of the carrier number $\gamma = 10^9$ s$^{-1}$, excess in the spin decay rate $\gamma_s = 50 \cdot 10^9$ s$^{-1}$, linear phase anisotropy $\gamma_p = 0$ s$^{-1}$, amplitude anisotropy $\gamma_a = 0$ s$^{-1}$.

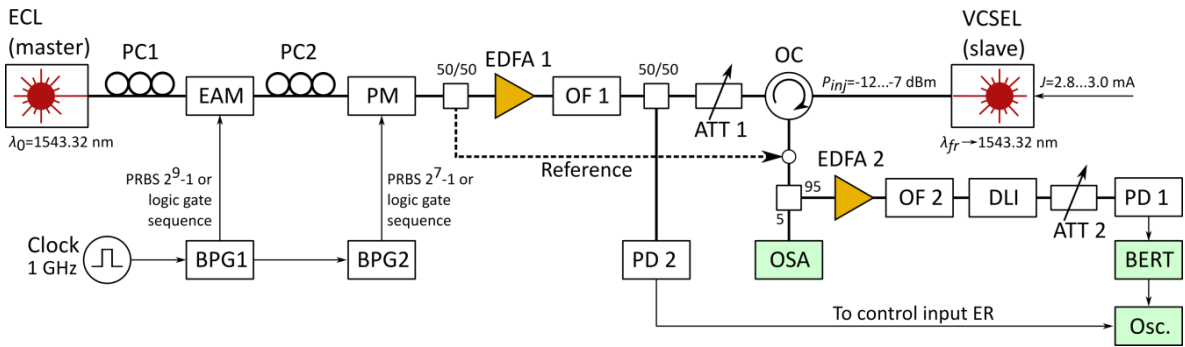

**Extended Data Figure 1 | Schematic illustration of the measurement setup.** Abbreviations: ECL – external cavity laser; PC – polarization controller; EAM – electro-absorption modulator; PM – phase modulator; BPG – bit pattern generator; EDFA – Erbium doped fibre amplifier; OF – optical filter; ATT – variable optical attenuator; OC – optical circulator; PD – p-i- n photodiode; OSA – optical spectrum analyzer; DLI – delay line interferometer; BERT – bit error rate tester; Osc. – digitizing oscilloscope.

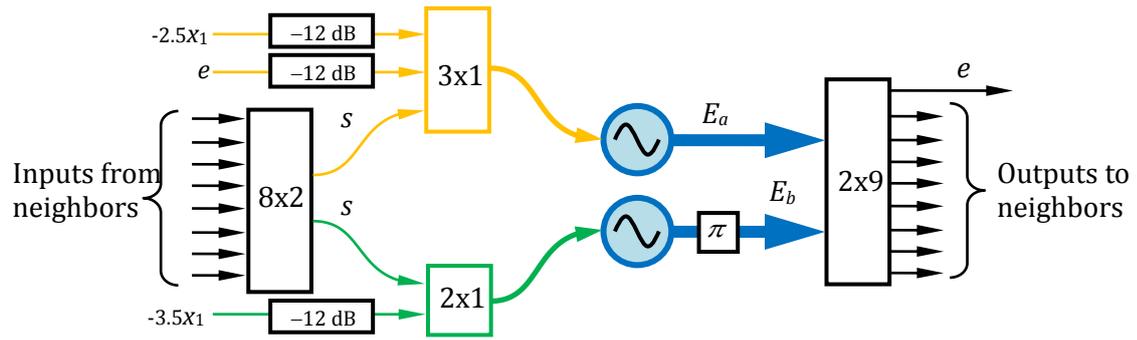

**Extended Data Figure 2 | Schematic illustration of a cell in Conway's Game of Life.**

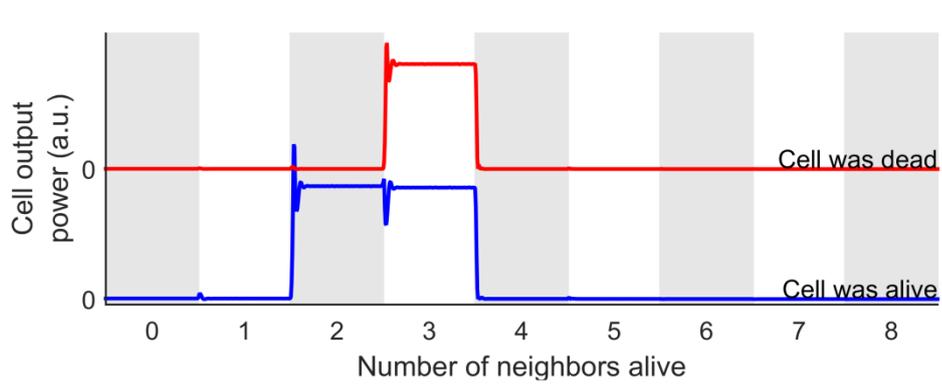

**Extended Data Figure 3 | Simulated output of the GoL cell with varying number of living neighbours.**

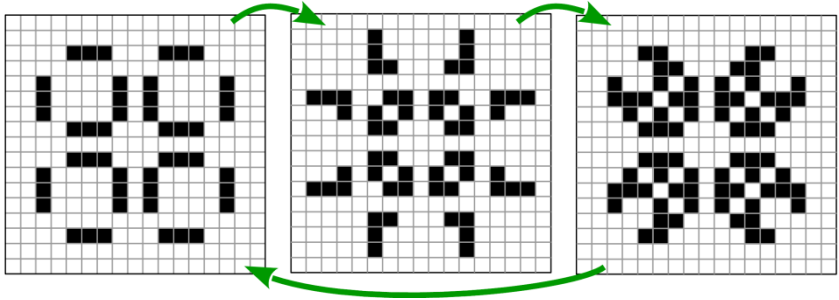

**Extended Data Figure 4 | A pulsating GoL pattern on a grid of 17x17 cells.**

# Supplementary Equations —
# 1. Steady-state analysis of an injection-locked laser
# 2. Universality of laser based neural networks
# 3. Simple constructions for OR and AND gates

Tuomo von Lerber,[1,2]* Matti Lassas,[2] Le Quang Trung,[1]
Vladimir Lyubopytov,[1,3] Arkadi Chipouline,[1] Klaus Hofmann,[4]
Franko Küppers[1,5]

[1]Photonics Lab, Technische Universität Darmstadt, Germany

[2]Department of Mathematics & Statistics, University of Helsinki, Finland

[3]Telecommunication Systems Dept., Ufa State Aviation Technical University, Russian Federation

[4]Integrated Electronic Systems Lab, Technische Universität Darmstadt, Germany

[5]College of Optical Sciences, University of Arizona, USA

*To whom correspondence should be addressed: tuomo.lerber@gmail.com.

March 21, 2017

## 1 Steady-state analysis of an injection-locked semiconductor laser

In following we will show that in steady-state an injection-locked laser can perform a two-dimensional complex normalization operation of linearly polarized electric fields. That is, the electric field of the emitted light adopts the master polarization and frequency, and locks the phase, while it regenerates the amplitude. This relation can be written as

$$\overline{E} = \overline{u} \, \|\overline{u}\|^{-1} e^{i\theta} \sqrt{\mu - 1}, \tag{1}$$

where $\mu$ is the pumping rate of the laser ($\mu = 1$ at the threshold), $\overline{E}, \overline{u} \in \mathbb{C}^2$ are the electric field amplitudes of the slave and the master, respectively. The angle $\theta$ is the phase shift between input and output signals. The carrier waves are taken to be of the form $\exp\left(-i\left(kz - \omega t\right)\right)$, that is, we use below the terminology that the electric field $E(t)\exp\left(-i\left(kz - \omega t\right)\right)$ has the slowly varying amplitude $E(t)$.

In absence of laser cavity anisotropies, the electric fields with slowly varying amplitudes $E_x(t)$ and $E_y(t)$ of the $x$ and $y$ polarized components, respectively, can be modeled with a set of rate equations [56]

$$\dot{E}_x = -\gamma_c E_x - i\gamma_c \alpha E_x + \gamma_c(1 + i\alpha)(NE_x + inE_y), \tag{2}$$

$$\dot{E}_y = -\gamma_c E_y - i\gamma_c \alpha E_y + \gamma_c(1 + i\alpha)(NE_y - inE_x), \tag{3}$$

$$\dot{N} = -\gamma\left[N\left(1 + |E_x|^2 + |E_y|^2\right) - \mu + in\left(E_y E_x^* - E_x E_y^*\right)\right], \tag{4}$$

$$\dot{n} = -\gamma_s n - \gamma\left[n\left(|E_x|^2 + |E_y|^2\right) + iN\left(E_y E_x^* - E_x E_y^*\right)\right], \tag{5}$$

where $N$ is a difference of normalized upper and lower level populations, $n$ is a normalized value of the difference between the population inversions, $\gamma_c$ is the decay rate of the electronic field in the cavity, i.e., $(2\gamma_c)^{-1}$ is the cavity photon lifetime; $\alpha$ is the linewidth enhancement factor, $\gamma$ is the decay rate of the total carrier number, and $\gamma_s$ is a decay rate that is related to electron angular momentum. In following we shall assume that the laser operates close to threshold $1 < \mu < 1.2$.

The master electric fields with the slowly varying amplitude $\overline{u}$ is a sum of input electric fields with the slowly varying amplitudes $\overline{e}_k$,

$$\overline{u} = \sum_k \overline{e}_j. \tag{6}$$

Note that we assume that the inputs $\overline{e}_k$ and the laser electric field $\overline{E}$ polarization components have the same carrier frequencies. Now the rate equations (2) and (3) can be written as

$$\dot{E}_x = -\gamma_c E_x - i\gamma_c \alpha E_x + \gamma_c(1+i\alpha)(NE_x + inE_y) + u_x, \tag{7}$$

$$\dot{E}_y = -\gamma_c E_y - i\gamma_c \alpha E_y + \gamma_c(1+i\alpha)(NE_y - inE_x) + u_y. \tag{8}$$

We assume a weak coupling $\|\bar{u}\| \ll \|\bar{E}\|$.

At the steady-state the time derivates of equations (4), (5), (7), and (8) will become zero

$$0 = -\gamma_c E_x - i\gamma_c \alpha E_x + \gamma_c(1+i\alpha)(NE_x + inE_y) + u_x, \tag{9}$$

$$0 = -\gamma_c E_y - i\gamma_c \alpha E_y + \gamma_c(1+i\alpha)(NE_y - inE_x) + u_y, \tag{10}$$

$$0 = -\gamma \left[ N \left(1 + \|E\|^2\right) - \mu + in \left(E_y E_x^* - E_x E_y^*\right) \right], \tag{11}$$

$$0 = -\gamma_s n - \gamma n \|E\|^2 + i\gamma N \left(E_y E_x^* - E_x E_y^*\right). \tag{12}$$

We can re-arrange the terms of equation (12) as

$$n = i\gamma N \frac{E_y E_x^* - E_x E_y^*}{\gamma_s + \gamma \|E\|^2}. \tag{13}$$

For linearly polarized field the difference $E_y E_x^* - E_x E_y^* = 0$ and thus $n = 0$ for steady-state. The remaining steady-state equations can be written as

$$0 = -\gamma_c E_x - i\gamma_c \alpha E_x + \gamma_c(1+i\alpha)NE_x + u_x, \tag{14}$$

$$0 = -\gamma_c E_y - i\gamma_c \alpha E_y + \gamma_c(1+i\alpha)NE_y + u_y, \tag{15}$$

$$0 = N \left(1 + \|E\|^2\right) - \mu, \tag{16}$$

and further simplified to

$$E_{x,y} = u_{x,y} \gamma_c^{-1}(1+i\alpha)^{-1}(1-N)^{-1}, \tag{17}$$

$$N = \mu \left(1 + \|E\|^2\right)^{-1}. \tag{18}$$

In order to solve equation (18) we proceed to calculate the electric field power from equation (17)

$$\|\overline{E}\|^2 = \|\overline{u}\|^2 \gamma_c^{-2}(1+\alpha^2)^{-1}(1-N)^{-2}, \tag{19}$$

and when substituted into equation (18)

$$N + N\|\overline{u}\|^2 \gamma_c^{-2}(1+\alpha^2)^{-1}(1-N)^{-2} - \mu = 0. \tag{20}$$

Further substitution of $N = 1 + \xi$ is inserted into equation (20) and the terms are rearranged as

$$\xi^3 + \xi^2(1-\mu) + \xi\|\overline{u}\|^2\gamma_c^{-2}(1+\alpha^2)^{-1} + \|\overline{u}\|^2\gamma_c^{-2}(1+\alpha^2)^{-1} = 0. \tag{21}$$

The cubic polynomial has a pair of roots in immediate vicinity of zero and we can approximate the equation with a second degree Taylor polynomial in neighborhood of $\xi = 0$ as

$$\xi^2(1-\mu) + \xi\|\overline{u}\|^2\gamma_c^{-2}(1+\alpha^2)^{-1} + \|\overline{u}\|^2\gamma_c^{-2}(1+\alpha^2)^{-1} = 0. \tag{22}$$

The equation (22) has roots

$$\xi^{(1)} = \frac{-\|\overline{u}\|^2\gamma_c^{-2}(1+\alpha^2)^{-1} + (\|\overline{u}\|^4\gamma_c^{-4}(1+\alpha^2)^{-2} - 4(1-\mu)\|\overline{u}\|^2\gamma_c^{-2}(1+\alpha^2)^{-1})^{1/2}}{2(1-\mu)}, \tag{23}$$

$$\xi^{(2)} = \frac{-\|\overline{u}\|^2\gamma_c^{-2}(1+\alpha^2)^{-1} - (\|\overline{u}\|^4\gamma_c^{-4}(1+\alpha^2)^{-2} - 4(1-\mu)\|\overline{u}\|^2\gamma_c^{-2}(1+\alpha^2)^{-1})^{1/2}}{2(1-\mu)} \tag{24}$$

and that we can write in the form $\Delta\xi \pm \xi$, that is, the roots are

$$\xi^{(1)} = \xi + \Delta\xi, \quad \xi^{(2)} = -\xi + \Delta\xi \tag{25}$$

where

$$\xi = \frac{(\|\overline{u}\|^4\gamma_c^{-4}(1+\alpha^2)^{-2} - 4(1-\mu)\|\overline{u}\|^2\gamma_c^{-2}(1+\alpha^2)^{-1})^{1/2}}{2(1-\mu)} \tag{26}$$

and $\Delta\xi > 0$.

A further simplification is achieved by noting that the cavity electric field decay rate $\gamma_c$ is typically large, for semiconductor lasers in range of $20 \cdot 10^{12} \text{s}^{-1}$ and thus $\gamma_c^{-2} = 2.5 \cdot 10^{-27} \text{s}^2$.

This implies that $\Delta\xi$ can be considered as a small parameter in the model and we can approximate

$$\xi = -\frac{\|\bar{u}\|}{\gamma_c\sqrt{1+\alpha^2}\sqrt{\mu-1}}. \tag{27}$$

We also define

$$N^{(1)} = 1 + \xi^{(1)}, \quad N^{(2)} = 1 + \xi^{(2)}. \tag{28}$$

If we consider left hand and right hand polarized electric fields with the slowly varying amplitudes

$$E_+ = \frac{1}{\sqrt{2}}(E_x + iE_y), \quad E_- = \frac{1}{\sqrt{2}}(E_x - iE_y),$$
$$u_+ = \frac{1}{\sqrt{2}}(u_x + iu_y), \quad u_- = \frac{1}{\sqrt{2}}(u_x - iu_y),$$

we see that (7) and (8) yield

$$\dot{E}_\pm = -\gamma_c E_\pm - i\gamma_c\alpha E_\pm + \gamma_c(1+i\alpha)(N \pm n)E_\pm + u_\pm, \tag{29}$$

that yields, with $n = 0$ and $N = N^{(k)}$, where $k = 1$ or $k = 2$,

$$\dot{E}_\pm = -(1+i\alpha)\gamma_c(1 - N^{(k)})E_\pm + u_\pm, \tag{30}$$

or

$$\frac{d}{dt}E_\pm = -\beta^{(k)}E_\pm + u_\pm, \tag{31}$$

where

$$\beta^{(k)} = (1+i\alpha)\gamma_c(1 - N^{(k)})$$
$$= -(1+i\alpha)\gamma_c\xi^{(k)}.$$

Similarly to (25), we write

$$\beta^{(1)} = \beta + \Delta\beta, \quad \beta^{(2)} = -\beta + \Delta\beta,$$

where

$$\beta = -(1+i\alpha)\gamma_c\xi, \quad \Delta\beta = -(1+i\alpha)\gamma_c\Delta\xi.$$

As $\xi < 0$ and $\Delta\xi$ is small, $\beta^{(1)}$ and $\beta^{(2)}$ are complex numbers with

$$\text{Re}\,\beta^{(1)} > 0, \quad \text{Re}\,\beta^{(2)} < 0.$$

The solution of (31) is given by

$$E_\pm(t) = \frac{1}{\beta^{(k)}}u_\pm + e^{-\beta^{(k)}t}\left(E_\pm(0) - \frac{1}{\beta^{(k)}}u_\pm\right), \quad t > 0. \tag{32}$$

From this we see that the ordinary differential equation (31), considered as dynamical system, has a fixed point $E_\pm = \frac{1}{\beta^{(k)}}u_\pm$. As $\text{Re}\,\beta^{(2)} < 0$, the solution corresponding to $k = 2$ is unstable and as $\text{Re}\,\beta^{(1)} > 0$, the solution corresponding to $k = 1$ is stable. Due to this we consider below the case $k = 1$. Then, we can approximate $\xi^{(1)}$ by

$$\xi^{(1)} \approx \xi = -\frac{\|\bar{u}\|}{\gamma_c\sqrt{1+\alpha^2}\sqrt{\mu-1}} \tag{33}$$

and thus,

$$N^{(1)} \approx N = 1 - \frac{\|\bar{u}\|}{\gamma_c\sqrt{1+\alpha^2}\sqrt{\mu-1}} \tag{34}$$

and when substituted into equation (17) we get

$$\begin{aligned}
E_x &= u_x\|\bar{u}\|^{-1}\sqrt{\mu-1}\exp\left[-i\text{Arg}\,(1+i\alpha)\right], \\
E_y &= u_y\|\bar{u}\|^{-1}\sqrt{\mu-1}\exp\left[-i\text{Arg}\,(1+i\alpha)\right].
\end{aligned} \tag{35}$$

For the purpose of this work the absolute phase $\theta = \text{Arg}\,(1+i\alpha)$ is of little interest, given that it stays constant in respect to the injected field. The amplitude multiplying coefficient $\sqrt{\mu-1}$ is important.

Summarising the above, we write the above relation in vectorial form as

$$\overline{E} = \bar{u}\|\bar{u}\|^{-1}e^{i\theta}\sqrt{\mu-1}, \tag{36}$$

which is an amplified normalization of complex vector $\bar{u} \in \mathbb{C}^2$. In other words, the injection-locked semiconductor laser adopts the linearly polarized master electric field and regenerates its amplitude.

## 2 Universality of laser based neural networks

In a suitable topology a collection of normalization operations can be used to approximate an arbitrary continuous function $f : \mathbb{C}^{j_0} \to \mathbb{C}^{\ell_0}$. We show this by using a laser based neutral network where the non-linearity of a node follows the normalization operation

$$\text{sign} : E \mapsto \|E\|^{-1} E \tag{37}$$

of the electric field, or a generalized normalization operator $\text{sign}_a$, and will present a proof for the uniform approximation theorem for such networks.

### 2.1 Mathematical model of a network of lasers

Let us consider a steady state in a neural network with a single non-linear layer as depicted in Fig. 1. In the first layer we denote the indexes of the nodes by $J = \{1, 2, \ldots, j_0\}$. The nodes in the first layer are not connected to each other and their electric field are denoted by $\mathbf{E}_j = (E_j^1, E_j^2) \in \mathbb{C}^2$, where $E_j^1$ corresponds to the vertical component of the electric field and $E_j^2$ corresponds to the horizontal component of the electric field. More precisely, the vertical and the horizontal electric fields are functions of time as $E_k^1 e^{i\omega t}$ and $E_k^2 e^{i\omega t}$, respectively.

In the second layer we denote the indexes of the nodes by $K = \{1, 2, \ldots, k_0\}$. In the node $N_k$ we have an electric field $e_k \in \mathbb{C}^2$ and an external source field $E_k^{ext} \in \mathbb{C}^2$. The nodes in the 2nd layer are connected to the 1st layer via matrices $A_{k,j} \in \mathbb{C}^{2\times 2}$ and to the external fields via matrices $B_k \in \mathbb{C}^{2\times 2}$.

In the 3rd layer we denote the indexes of the nodes by $L = \{1, 2, \ldots, \ell_0\}$. In these nodes we have electric fields $\mathcal{E}_\ell \in \mathbb{C}^2$. The nodes in the 3rd layer are connected to the nodes in the 2nd layer via matrixes $C_{\ell,k} \in \mathbb{C}^{2 \times 2}$.

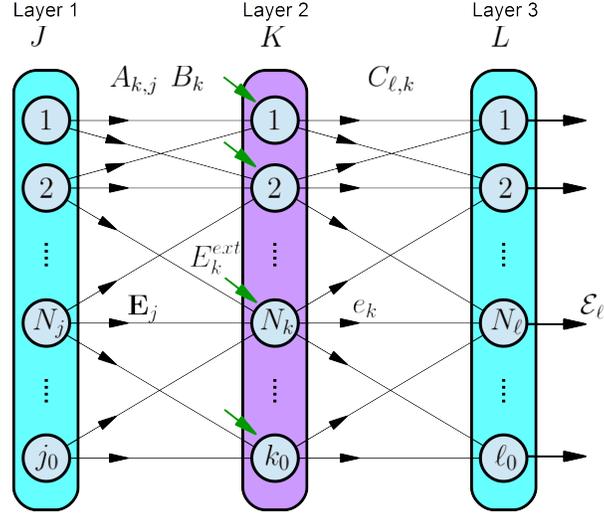

Figure 1: Neural network with a non-linear middle layer.

For the above objects we consider a model

$$u_k = (\sum_{j \in J} A_{k,j} \mathbf{E}_j) + B_k E_k^{ext}, \quad k \in K, \tag{38}$$

$$\widehat{u}_k = \frac{1}{\|u_k\|_{\mathbb{C}^2}} u_k, \quad k \in K,$$

$$e_k = c_0 \widehat{u}_k, \quad c_0 \in \mathbb{C},$$

$$\mathcal{E}_\ell = \sum_{k \in K} C_{\ell,k} e_k, \quad \ell \in J,$$

where $c_0$ is a constant that determines the emission amplitude of the laser $N_k$. This model defines function $\mathcal{M} : (\mathbb{C}^2)^{j_0} \to (\mathbb{C}^2)^{\ell_0}$,

$$\mathcal{M} : (\mathbf{E}_j)_{j=1}^{j_0} \mapsto (\mathcal{E}_\ell)_{\ell=1}^{\ell_0}. \tag{39}$$

Note that we could add more connections in the above network. The model would correspond then to a case where matrices modeling such connections are close to zero.

To consider the use of a laser network as a computational unit, we will consider the input signals $(\mathbf{E}_j)_{j=1}^{j_0}$ where the all $\mathbf{E}_j$ have vanishing horizontal component. Then, to simplify notations, we will replace the complex vector valued signal $(\mathbf{E}_j)_{j=1}^{j_0}$, consisting of $j_0$ complex numbers and $j_0$ zeros, by complex vector valued signal $x = (x_k)_{k=1}^{j_0} \in \mathbb{C}^{j_0}$, consisting of $j_0$ complex numbers. Also, we assume that the matrices $A_{k,j}$ and $C_{\ell,k}$ map the vertically polarised signal (i.e., the signals with vanishing horizontal component) to vertically polarised signals and horizontally polarised signals to zero, that is

$$A_{k,j} = \begin{pmatrix} A_{k,j}^{11} & 0 \\ 0 & 0 \end{pmatrix}, \quad C_{\ell,k} = \begin{pmatrix} C_{\ell,k}^{11} & 0 \\ 0 & 0 \end{pmatrix}. \tag{40}$$

This is the same than equipping the optical paths of the matrices $A_{j,k}$ and $C_{\ell,k}$ with a polarizing filter $F$ that passes the vertical and removes the horizontal polarization components.

We will consider a theoretical device that correspond to a continuous function $f : \mathbb{C}^{j_0} \to \mathbb{C}^{\ell_0}$, that maps the input $\mathbf{E} \in \mathbb{C}^{j_0}$ to the output $f(\mathbf{E}) \in \mathbb{C}^{\ell_0}$, and approximate it with connected lasers modelled by function $\widetilde{f}$. Indeed, we assume that the input signal satisfies the bound $|x| < R$ and show that when the laser based neural network is large enough, $\widetilde{f}(x)$ approximates $f(x)$ with an arbitrary given precision.

Let us explain the idea of this result in non-rigorous terms in the case when $m = 1$ and when the input signal is a real valued, that is, the slowly varying amplitudes are given by a real vector $x \in \mathbb{R}^{j_0}$. This is done by considering so-called ridge functions $\psi(x \cdot v + c)$, where $v \in \mathbb{R}^{j_0}$ and $c \in \mathbb{R}$. The ridge function combines a dimension reducing operation $x \mapsto x \cdot v + c$ that maps $n$-dimensional vector to a (1-dimensional) real number. This 1-dimensional signal is processed with a non-linear function $\psi : \mathbb{R} \to \mathbb{R}$ that is sufficiently regular and is not a polynomial. Using approximation theory, see [51, 55, 66], we have that function $f(x)$ can be approximated in the set $|x| < R$ by a sum of ridge functions, that is,

$$f(x) \approx a_1 \psi(x \cdot v_1 + c_1) + a_2 \psi(x \cdot v_2 + c_2) + \cdots + a_K \psi(x \cdot v_K + c_K),$$

where $v_1, v_2, \ldots, v_K$ are vectors in $\mathbb{R}^n$ and $a_1, a_2, \ldots, a_K$ and $c_1, c_2, \ldots, c_K$ are real numbers. In particular, $\psi$ can be a sign-operation that can be written in terms of a normalisation operation,

$$\text{sign}(s) = \frac{s}{|s|}, \tag{41}$$

that can be implemented via a steady state of an injection-locked laser. The vectors $v_k$ correspond to the matrices $(A_{j,k})_{j=1}^{j_0}$. The numbers $c_k$ are determined by $B_k E_k^{ext}$, that is, by the external fields.

Note that in the above simplified explanation we analysed only inputs $x \in \mathbb{R}^{j_0}$ that are real valued vectors. As the amplitudes of electric fields are complex, we need to consider complex valued input vectors and in following we will formulate and prove a theorem on approximation of functions $f : \mathbb{C}^{j_0} \to \mathbb{C}^{\ell_0}$ by a linear combination of complex sign-functions.

## 2.2 Generalized sign operators

One may note that the sign-function is unstable at the zero, where even a small perturbation of the input could flip the sign of the output. To obtain stable processing we set[1] the incoming signals $\mathbf{E}_j$ to be vertically polarised and the constant external signals (the bias) $E_k^{ext}$ to have both vertical and horizontal polarization components. Here we assume that the lasers have no polarization anisotropy, i.e., the emission or locking susceptibility have no preference in terms of polarization. Under these circumstances the normalization operator can be replaced with a smoothed normalization operator (i.e., a generalized complex sign function)

$$\text{sign}_a(s) = \frac{s}{(|s|^2 + a^2)^{1/2}}, \tag{42}$$

where the matrix $B_k$ in Figure 1 is the identity operator, $a = |B_k E_k^{2,ext}| = |E_k^{2,ext}|$ is the amplitude of the horizontal component of bias signal $E_k^{ext}$, and $s$ is the sum of the vertically

---

[1] Applies only this mathematical analysis. In the experimental setup both the input signal and the bias had the same state of polarization

polarised input signal $E_k$ and the vertical component $E_k^{1,ext}$ of bias signal $E_k^{ext}$. Here we allow $a$ to have values $a \geq 0$ such that also the normalisation operator can be considered as a special case of the smoothed normalisation operator when the parameter $a$ takes value of zero. Also, the function sign, defined in (41), extends to a function $\text{sign} : \mathbb{C} \setminus \{0\} \to \mathbb{C}$, and and $\text{sign}_a$, defined in (42) with $a > 0$, extends to a function $\text{sign}_a : \mathbb{C} \to \mathbb{C}$, that is,

$$\text{sign}(z) = \frac{z}{|z|}, \qquad \text{sign}_a(z) = \frac{z}{(|z|^2 + a^2)^{1/2}}. \tag{43}$$

To make the above more specific, assume that $E_k^{ext} = (E_k^{1,ext}, E_k^{2,ext})$ where $E_k^{1,ext} \in \mathbb{C}$ and $E_k^{2,ext} \in \mathbb{C}$ are the horizontal and the vertical components of the bias electric field, respectively. Also, assume that $\mathbf{E}_j$ have vanishing horizontal components, the optical paths corresponding the matrices $A_{j,k}$ and $C_{\ell,k}$ are equipped with a polarizing filter $F$, that is, matrices $A_{j,k}$ and $C_{\ell,k}$ have the form (40). Finally, assume that $B_k$ are identity matrixes and $|E_k^{2,ext}| = a$ for all $k$. Then above considered model (38)-(39) of a laser network, can thus be refined using functions (42) as follows:

Let

$$\begin{aligned} U_k &= (\sum_{j \in J} A_{k,j}^{11} \mathbf{E}_j) + E_k^{1,ext} \quad k \in K, \\ \widehat{U}_k &= \text{sign}_a(U_k) \quad k \in K, \\ E_k &= c_0 \widehat{U}_k, \\ \mathcal{E}_\ell &= \sum_{k \in K} C_{\ell,k}^{11} E_k, \quad \ell \in J, \end{aligned} \tag{44}$$

where $a > 0$ is small and $A_{k,j}^{11}$ and $C_{k,j}^{11}$ are complex numbers and $c_0 \in \mathbb{C}$ is a constant. This refined model defines function $\mathcal{M}_a : \mathbb{C}^{j_0} \to \mathbb{C}^{\ell_0}$,

$$\mathcal{M}_a : (\mathbf{E}_j)_{j=1}^{j_0} \mapsto (\mathcal{E}_\ell)_{\ell=1}^{\ell_0}. \tag{45}$$

## 2.3 The approximation theorem

Our main approximation result for functions $f : \mathbb{C}^{j_0} \to \mathbb{C}^{\ell_0}$ in this section is the following:

**Theorem 1.** *Let $j_0, \ell_0 \in \mathbb{Z}_+$ and $a \geq 0$. Then for any continuous functions $f : \mathbb{C}^{j_0} \to \mathbb{C}^{\ell_0}$ and $R > 0$ and $\varepsilon > 0$, there are $A^{11}_{k,j}$, $C^{11}_{k,j}$ and $E^{1,ext}_k$, defining function $\mathcal{M}_a$, such that*

$$\left\| f((\mathbf{E}_j)_{j=1}^{j_0}) - \mathcal{M}_a((\mathbf{E}_j)_{j=1}^{j_0}) \right\|_{\mathbb{C}^{\ell_0}} < \varepsilon_1, \quad \textit{for all } z \in \mathbb{C}^{j_0} \textit{ satisfying } |z| < R. \tag{46}$$

Note that Theorem 1 is different to the standard approximation theorems concerning real valued signals, that are used e.g. in theory of neural networks, in the sense that the generalized complex sign functions $\text{sign}_a : \mathbb{C} \to \mathbb{R}$ are quite different to the sigmoid type functions used in the real approximation theory.

We will prove Theorem 1 below after we have introduced two other results – first, the approximation of ridge functions by a sum of generalized complex sign-functions defined in the complex plane, and second, approximation of arbitrary functions by ridge functions. To make the proof readable for a wider audience, we have included detailed explanations of the used mathematical constructions, as well as some motivating discussions.

*2.3.1 Approximation of a function of one complex variable by a sum of $\text{sign}_a$ functions*

We denote by $B_{\mathbb{C}}(0, r_0) = \{z \in \mathbb{C};\ |z| < r_0\}$ the open disc in $\mathbb{C}$ with centre zero and radius $r_0$ and by $\overline{B}_{\mathbb{C}}(0, r_0) = \{z \in \mathbb{C};\ |z| \leq r_0\}$ the corresponding disc containing the boundary circle.

Next we formulate and prove an auxiliary result on the approximation of a function of one complex variable by a sum of generalized complex sign functions.

**Lemma 1.** *For any $a \geq 0$, $r_0 > 0$, a continuous function $g : \overline{B}_{\mathbb{C}}(0, r_0) \to \mathbb{C}$ and $\varepsilon_0 > 0$ there are $i_0 \in \mathbb{Z}_+$ and $b_i \in \mathbb{C}$, and $z_i \in \mathbb{C}$, $i = 1, 2, \ldots, i_0$, such that*

$$\left| g(z) - \sum_{i=1}^{i_0} b_i \text{sign}_a (z - z_i) \right| < \varepsilon_0 \quad \text{for all } |z| \leq r_0. \tag{47}$$

*In other words, the linear combinations of the functions $\{\text{sign}_a (z - z'); \ z' \in \mathbb{C}\}$ is a dense subset in the space of continuous, complex valued functions $C(\overline{B}_{\mathbb{C}}(0, r_0))$.*

**Proof.** Next we identify $\mathbb{C}$ with $\mathbb{R}^2$, that is, the complex number $x_1 + ix_2 \in \mathbb{C}$ is identified with $(x_1, x_2) \in \mathbb{R}^2$. We denote by $B_{\mathbb{R}^2}(0, r_0)$ the disc in $\mathbb{R}^2$ with centre zero and radius $r_0$. Without loss of generality, we assume below that $0 < \varepsilon_0 < \frac{1}{4}$. Below, we use also $s > 0$.

Let $h_a(z) = \text{sign}_a(z)$ and $h_{a,z_0}(z) = h_a(z - z_0)$. Also, let

$$h_a^s(z) = h_a * \phi_s(z) = \int_{\mathbb{R}^2} h_a(z') \phi_s(z - z') \, dz'$$

be the convolution of $h_a$ and $\phi_s$ and denote $h_{a,z_0}^s(z) = h_a^s(z - z_0)$. Here, $\phi_s(x) = s^{-2}\phi(x/s)$ is a mollifier, that is, a non-negative function $\phi \in C^\infty(\mathbb{R}^2)$ that vanishes outside the disc $B_{\mathbb{R}^2}(0, 1)$ of radius 1 and satisfies $\int_{\mathbb{R}^2} \phi(x) dx = 1$. We will also assume that $|\phi(x)| \leq 1$, implying that $|\phi_s(x)| \leq s^{-2}$. Recall that a function $h(x)$, $x \in \mathbb{R}^2$ is real-analytic if for every $x_0 \in \mathbb{R}^2$ there is $\rho > 0$ such that the Taylor series of $h(x)$, computed with the centre $x_0$, converges to $h(x)$ for every $x \in B_{\mathbb{R}^2}(x_0, \rho)$. As $\phi_s(x)$ is zero outside a bounded disc, the Fourier transform $\widehat{\phi}_s(\xi)$ is a real-analytic function that does not vanish in any open set [58, Theorem 7.3 and the proof of Lemma 7.2].

Since $(|x|^2 + a^2)^{1/2} h_a(x) = x_1 + ix_2$, we have for the Fourier transform

$$(-\nabla_\xi^2 + a^2)^{1/2} \widehat{h}_a(\xi) = -i\left(\frac{\partial}{\partial \xi_1} + i\frac{\partial}{\partial \xi_2}\right) \delta_0(\xi) \tag{48}$$

where $\delta_0(\xi)$ is Dirac's delta distribution supported at zero.

Next we show that $\widehat{h}_a(\xi)$ does not vanish in an open subset of $\mathbb{R}^2 \setminus \{0\}$. To show this, assume the opposite, that $\widehat{h}_a(\xi)$ vanishes in an open set of $\mathbb{R}^2 \setminus \{0\}$. The inverse operator $((-\nabla_\xi^2 + a^2)^{1/2})^{-1}$, is a so-called analytic pseudo-differential operator, in sense of [63, Section 5.2, formulas (2.2)-(2.4)], with amplitude $k(\eta) = (|\eta|^2 + a^2)^{-1/2}$, and the right hand side of (48) is real-analytic in the set $\mathbb{R}^2 \setminus \{0\}$. By [63, Section 5.2, Thm 2.1], such operators map real-analytic functions to real-analytic functions. This results and the equation (48) yield that also $\widehat{h}_a(\xi)$ is real-analytic in the set of $\mathbb{R}^2 \setminus \{0\}$. Hence, if $\widehat{h}_a(\xi)$ vanishes in an open subset of $\mathbb{R}^2 \setminus \{0\}$, then by [58, Theorem 7.3 and the proof of Lemma 7.2], $\widehat{h}_a(\xi)$ has to vanish in the whole set $\mathbb{R}^2 \setminus \{0\}$. Then by [58, Theorem 6.25], $\widehat{h}_a(\xi)$ is a finite sum of delta-distributions at zero and its derivatives, thus $h_a(x)$ is a polynomial.

As function $h_a(x)$, given in (42), is not a polynomial we have obtained a contradiction. This implies that $\widehat{h}_a(\xi)$ can not vanish in an open set of $\mathbb{R}^2 \setminus \{0\}$. Let $r_0 > 0$ and consider the space $L^2(B_{\mathbb{R}^2}(0, r_0))$ of square integrable functions $B_{\mathbb{R}^2}(0, r_0) \to \mathbb{C}$. Let $Y_{a,s} \subset L^2(B_{\mathbb{R}^2}(0, r_0))$ be the linear subspace, spanned by the functions $h_{a,z}^s$, where $z \in \mathbb{C}$. To show that $Y_{a,s}$ is dense in $L^2(B_{\mathbb{R}^2}(0, r_0))$, assume that $f \in L^2(B_{\mathbb{R}^2}(0, r_0))$ is a function satisfying

$$\int_{B_{\mathbb{R}^2}(0,r_0)} f(x)p(x)\,dx = 0, \quad \text{for all } p \in Y_{a,s}. \tag{49}$$

By [57, Theorem 4.11], to show that $Y_{a,s}$ is a dense subspace it is enough to show that the function $f$ satisfying (49) is zero. Let $F \in L^2(\mathbb{R}^2)$ be the zero continuation of the function $f$, that is, $F(x)$ is equal to $f(x)$ in $B_{\mathbb{R}^2}(0, r_0)$ and zero elsewhere.

Then by (49),

$$\int_{\mathbb{R}^2} F(x)h_a^s(x-z)\,dx = 0, \quad \text{for all } z \in \mathbb{C}. \tag{50}$$

Hence, the convolution $F * h_a^s(z)$ is zero for all $z \in \mathbb{C}$.

The Fourier-transform of $F * h_a^s(z)$ satisfies

$$\widehat{F * h_a^s}(\xi) = \widehat{F}(\xi)\,\widehat{h_a^s}(\xi) = \widehat{F}(\xi)\,\widehat{h_a}(\xi)\,\widehat{\phi_s}(\xi) = 0, \quad \text{for all } \xi \in \mathbb{R}^2. \quad (51)$$

Since $\widehat{h_a}(\xi)$ and $\widehat{\phi_s}(\xi)$ do not vanish in any open set, we see that $\widehat{F}(\xi)$ is zero almost everywhere, and hence $F = 0$. Thus $f = 0$, implying that $Y_{a,s}$ is dense in $L^2(B_{\mathbb{R}^2}(0, r_0))$.

Next we improve the above argument by changing the function space $L^2$ to a Sobolev space. Let us consider $Y_{a,s}$ as a subset of the Sobolev space $H^k(B_{\mathbb{R}^2}(0, r_0))$ with $k \geq 2$, see [44]. The norm of the Sobolev space $H^k(B_{\mathbb{R}^2}(0, r_0))$ is given by

$$\|f\|_{H^k(B_{\mathbb{R}^2}(0,r_0))} = \Big( \sum_{j=0}^{k} \int_{B_{\mathbb{R}^2}(0,r_0)} |\nabla^j f(x)|^2 \, dx \Big)^{1/2}.$$

Assume that $f$ is in the dual space of $H^k(B_{\mathbb{R}^2}(0, r_0))$ and satisfies (49). By [64, Thm. 4.8.1], the dual space of the Hilbert space $H^k(B_{\mathbb{R}^2}(0, r_0))$ can be identified with the space $\widetilde{H}^{-k}(B_{\mathbb{R}^2}(0, r_0))$ that consists of functions in the Sobolev space $H^{-k}(\mathbb{R}^2)$ that vanish outside $\overline{B}_{\mathbb{R}^2}(0, r_0)$.

The equation (49) implies that the zero continuation $F \in H^{-k}(\mathbb{R}^2)$ of $f$ satisfies (50). Again, this yields that the Fourier transform $\widehat{F}(\xi)$ satisfies (51) and hence $F = 0$. This implies $Y_{a,s}$ is dense in also in the space $H^k(B_{\mathbb{R}^2}(0, r_0))$. This, the density of $C^\infty$-smooth functions in the space of continuous function $C(\overline{B}_{\mathbb{R}^2}(0, r_0))$ and the Sobolev embedding theorem, see [44], imply that $Y_{a,s}$ is dense also in the space $C(\overline{B}_{\mathbb{R}^2}(0, r_0))$. Thus for any $g \in C(\overline{B}_{\mathbb{R}^2}(0, r_0))$ and $r_0 > 0$ and $\varepsilon_0 > 0$ there are $i_0 \in \mathbb{Z}_+$, $b_i \in \mathbb{C}$ and $z_i' \in \mathbb{R}^2$ such that

$$\Big| g(z) - \sum_{i=1}^{i_0} b_i h_a^s(z - z_i') \Big| < \varepsilon_0/2 \quad \text{for all } |z| \leq r_0. \quad (52)$$

Now we will approximate the convolution $h_a^s(z) = h_a * \phi_s$, at a point $z \in \mathbb{R}^2$, by a Riemann sum: Let $\widetilde{\varepsilon}_0 = \varepsilon_0/C_0$, where the number $C_0 \geq 1$ is determined later and $\eta = \widetilde{\varepsilon}_0^2$. Note that the function $\phi_s(x)$ vanishes outside the disc of radius $s$. Let us then enumerate the points in the set $\{(\eta k_1, \eta k_2) \in B_{\mathbb{R}^2}(0, 2(s+1));\ (k_1, k_2) \in \mathbb{Z}^2\}$ as $z_j'' \in \mathbb{R}^2$, $j = 1, 2, \ldots, J$. Moreover, let

$$U_j = \{(x,y) \in \mathbb{R}^2;\ |x - \operatorname{Re} z_j''| \leq \eta/2,\ |y - \operatorname{Im} z_j''| \leq \eta/2\}$$

be the squares centered at the points $z_j''$. Assume that there is $j_0(\eta) \in \mathbb{Z}_+$ such that $|z_j'' - z| \leq \eta^{1/2}$ for $j = 1, 2, \ldots, j_0(\eta)$ and that $|z_j'' - z| > \eta^{1/2}$ for $j > j_0(\eta)$. As the squares $U_j$, $j \leq j_0(\eta)$ are contained in the disc $B_{\mathbb{R}^2}(z, 2\eta^{1/2})$, we see that $j_0(\eta) < C_1(\eta^{1/2})^2/\eta^2 = C_1\eta^{-1}$, where $C_1 = 4\pi$. Then we will approximate $h_a^s = h_a * \phi_s$ by a Riemann sum

$$h_a * \phi_s(z) = \int_{\mathbb{R}^2} \phi_s(z'')\, h_a(z - z'')dz'' \approx \sum_{j=1}^{J} \Phi_{s,j}\, h_a(z - z_j'')\eta^2,$$

where we denote

$$\Phi_{s,j} = \frac{1}{\eta^2} \int_{U_j} \phi_s(z'')dz''.$$

The error in the above approximation $\approx$ will be estimated below more precisely. To that end, observe that

$$\Big| h_a * \phi_s(z) - \sum_{j=1}^{J} \Phi_{s,j} h_a(z - z_j'')\eta^2 \Big| \leq I_1 + I_2, \quad \text{where}$$

$$I_1 = \sum_{j=1}^{j_0(\eta)} \int_{U_j} \phi_s(z'')|h_a(z - z'') - h_a(z - z_j'')|dz'',$$

$$I_2 = \sum_{j=j_0(\eta)+1}^{J} \int_{U_j} \phi_s(z'')|h_a(z - z'') - h_a(z - z_j'')|dz''.$$

Note that as $|h_a| \leq 1$, we have

$$|I_1| \leq s^{-2}\eta^2 \cdot j_0(\eta) \leq C_1 s^{-2}\eta \leq C_1 s^{-2}\widetilde{\varepsilon}_0.$$

For $j > j_0(\eta)$, we have $|z_j'' - z| > \eta^{1/2}$. Then for $z'' \in U_j$ we have

$$|z'' - z_j''| \leq 2\eta \leq \frac{1}{2}\eta^{1/2},$$

implying $|z'' - z| > \frac{1}{2}\eta^{1/2}$ and

$$\|\nabla_{z''} h_a(z - z'')\|_{\mathbb{R}^2} \leq |z'' - z_j''|^{-1} \leq 2\eta^{-1/2}. \qquad (53)$$

Moreover, for $z'' \in U_j$ we have $|z'' - z_j''| < 2\eta$ and hence (53) implies for $z \in U_j$

$$\begin{aligned}|h_a(z - z'') - h_a(z - z_j'')| &\leq (\max_{y \in U_j}\|\nabla h_a(z - y)\|_{\mathbb{R}^2})|z'' - z_j''| \\ &\leq 4\eta^{1/2} = 4\widetilde{\varepsilon}_0.\end{aligned}$$

Since $\phi_s(x) \geq 0$ and $\int_{\mathbb{R}^2} \phi_s(x)dx = 1$, this implies

$$|I_2| \leq 4\eta^{1/2} = 4\widetilde{\varepsilon}_0.$$

This shows that there are $c_j = \Phi_{s,j}\eta^2 \in \mathbb{C}$ and $z_j'' \in \mathbb{R}^2$ such that

$$\left|h_a^s(z) - \sum_{j=1}^J c_j h_a(z - z_j'')\right| < (C_1 s^{-2} + 4)\widetilde{\varepsilon}_0 \quad \text{for all } z \in \mathbb{C}. \qquad (54)$$

Now, let us choose $C_0$, that was above used to define $\widetilde{\varepsilon}_0$ so that $\widetilde{\varepsilon}_0 = \varepsilon_1/C_0$, to be

$$C_0 = (C_1 s^{-2} + 4)2i_0(1 + \max_i |b_i|).$$

Then, (54) yields

$$\left|\sum_{i=1}^{i_0} b_i h_a^s(z - z_i') - \sum_{i=1}^{i_0}\sum_{j=1}^J b_i c_j\, h_a(z - z_j'')\right| < \frac{\varepsilon_0}{2}, \quad \text{for all } |z| \leq r_0. \qquad (55)$$

By combining (52) and (55), we obtain the claim of the lemma. $\qquad\square$

*2.3.2 Approximation of functions in $\mathbb{R}^n$ by ridge functions*

Next we recall some results of approximation theory for functions $f : \mathbb{R}^n \to \mathbb{R}^m$. A function $\psi : \mathbb{R} \to \mathbb{R}$ is a piecewise continuous function if there are $s_1, s_2, \ldots, s_k \in \mathbb{R}$, such that $s_j < s_{j+1}$ and the restrictions of the function $\psi(s)$ are continuous on open intervals $(-\infty, s_1)$, $(s_1, s_2)$, $(s_2, s_3)$,..., $(s_{k-1}, s_k)$, and $(s_k, \infty)$. Also, the function $\psi : \mathbb{R} \to \mathbb{R}$ is bounded if there is $M > 0$ such that $|\psi(s)| \leq M$ for all $s \in \mathbb{R}$.

Next we will use the following result of approximation theory functions in $\mathbb{R}^n$, see [51], [55] and also [66].

**Theorem 2.** *Let $n, m \in \mathbb{Z}_+$ and $\psi : \mathbb{R} \to \mathbb{R}$ be a bounded and piecewise continuous function that is not a polynomial. Then for any continuous functions $f : \mathbb{R}^n \to \mathbb{R}^m$ and $r_1 > 0$ and $\varepsilon_1 > 0$, there are $K \in \mathbb{Z}_+$, $v_k \in \mathbb{R}^n$, $d_k \in \mathbb{R}^m$, and $c_k \in \mathbb{R}$, $k = 1, 2, \ldots, K$ such that*

$$\left\| f(x) - \sum_{k=1}^{K} d_k \psi(x \cdot v_k - c_k) \right\|_{\mathbb{R}^m} < \varepsilon_1, \quad \textit{for all } x \in \mathbb{R}^n \textit{ satisfying } |x| < r_1. \tag{56}$$

*2.3.3 Proof of the approximation theorem*

By combining the above results, we can now prove the approximation result for laser based systems:

> **Proof of Theorem 1.** Below, we identify $z = (z_j)_{j=1}^{j_0} \in \mathbb{C}^{j_0}$ with $x = (x_k)_{k=1}^{2j_0} \in \mathbb{R}^{2j_0}$, where $z_j = x_{2j-1} + ix_{2j}$. In such manner, we can identify $\mathbb{C}^{j_0}$ with $\mathbb{R}^n$, $n = 2j_0$ and $\mathbb{C}^{\ell_0}$ with $\mathbb{R}^m$, $m = 2\ell_0$.
>
> Let $f : \mathbb{C}^{j_0} \to \mathbb{C}^{\ell_0}$ be a continuous function that can be identified with a function $f : \mathbb{R}^n \to \mathbb{R}^m$. Also, let $\varepsilon > 0$ and $R > 0$.
>
> Let $r_1 = R$, $\varepsilon_1 = \varepsilon/2$ and use a $C^\infty$-smooth function $\psi(y)$ that is not polynomial, e.g. $\psi(y)$ is $\sin(y)$. By Theorem 2, there are $K \in \mathbb{Z}_+$, $v_k \in \mathbb{R}^n$, $d_k \in \mathbb{R}^m$, and $c_k \in \mathbb{R}$, $k = 1, 2, \ldots, K$ such that (56) is valid.

Then for $(v_k)_{k=1}^K$, $v_k \in \mathbb{R}^{2j_0}$ we define $w^{(k)} = v_{2k-1} - iv_{2k} \in \mathbb{C}^{j_0}$, and denote $w^{(k)} = (w_j^{(k)})_{j=1}^{j_0} \in \mathbb{C}^{j_0}$. With these notations, we can write

$$x \cdot v_k = \operatorname{Re} \big( \sum_{j=1}^{j_0} z_j w_j^{(k)} \big), \tag{57}$$

so that

$$\psi(x \cdot v_k - c_k) = g((\sum_{j=1}^{j_0} z_j w_j^{(k)}) - c_k), \quad g(z) = \psi(\operatorname{Re} z). \tag{58}$$

Then by Lemma 1, for the function $g(z)$ defined above and for

$$r_0 = 2j_0(\max_k |v_k| R + \max_k |c_k| + 1),$$
$$\varepsilon_0 = \varepsilon/(2K \max_j(1 + |d_j|)),$$

there are $i_0 \in \mathbb{Z}_+$ and $b_i \in \mathbb{C}$ and $z_i' \in \mathbb{C}$ such that

$$\Big| g(z) - \sum_{i=1}^{i_0} b_i \operatorname{sign}_a (z - z_i') \Big| < \varepsilon/2 \quad \text{for all } z \in \mathbb{C} \text{ such that } |z| \le r_0. \tag{59}$$

Combining the inequality (56) and (59), where $g$ is given by (58), we obtain

$$\big\| f(x) - \mathcal{M}_a(x) \big\|_{\mathbb{R}^m} < \varepsilon, \quad \text{for all } x \in \mathbb{R}^n, |x| < R, \text{ where} \tag{60}$$

$$\mathcal{M}_a(x) = \sum_{k=1}^K \sum_{i=1}^{i_0} d_k b_i \operatorname{sign}_a \Big( (\sum_{j=1}^{j_0} z_j w_j^{(k)}) - c^{(k)} - z_i' \Big). \tag{61}$$

As $\mathcal{M}_a(x)$ is of the required form (44), the claim is proven. $\square$

To end this section, let us explain shortly a constructive method to obtain the above approximations.

Below, we use a non-polynomial function $g(z) = \operatorname{sign}_a (\operatorname{Re} z)$ where $a > 0$ is small. First we consider, how Lemma 1 on the approximation of a continuous function $g(z)$, $z \in \mathbb{C}$, by a sum of smoothed normalisation functions, can be done in practice. To this end, we use Tikhonov regularisation [46]. This means solving the minimisation problem

$$\min \left( \left\| g(z) - \sum_{i=1}^{i_0} b_i \mathrm{sign}_a \left( z - z'_i \right) \right\|^2_{H^2(B_{\mathbb{R}^2}(0,r_0))} + \sum_{i=1}^{i_0} \gamma |b_i|^2 \right) \quad (62)$$

where minimisation is taken over $b = (b_i)_{i=1}^{i_0} \in \mathbb{C}^{i_0}$. Here, $\gamma > 0$ is a small parameter, called the regularisation parameter and $z'_1, z'_2, \ldots, z'_{i_0} \in \mathbb{R}^2$ are a computational grid in a disc $B_{\mathbb{R}^2}(0, r_1)$ with some $r_1 > 0$. The solution of this quadratic minimisation problem can be written as

$$b = (A + \gamma I)^{-1} w, \quad (63)$$

where matrices $A = [A_{ij}]_{i,j=1}^{i_0}$ and $w = [w_j]_{j=1}^{i_0}$ are given by

$$\begin{aligned} A_{ij} &= \int_{B_{\mathbb{R}^2}(0,r_0)} (1-\Delta)\psi_i(x) \overline{(1-\Delta)\psi_j(x)} \, dx, \\ w_j &= \int_{B_{\mathbb{R}^2}(0,r_0)} (1-\Delta)g(x) \overline{(1-\Delta)\psi_j(x)} \, dx \end{aligned}$$

and $\psi_i(x) = \mathrm{sign}_a(x - z_i)$. Here, when $x = (x_1, x_2) \in \mathbb{R}^2$, we have denoted $\Delta = \frac{\partial^2}{\partial x_1^2} + \frac{\partial^2}{\partial x_2^2}$.

The above gives an algorithmic way to obtain an approximation

$$\mathrm{sign}_a(\mathrm{Re}\, z) \approx \sum_{i=1}^{i_0} b_i \mathrm{sign}_a(z - z_i), \quad \text{for } z \in B_{\mathbb{C}}(0, r_0), \quad (64)$$

that corresponds to Lemma 1.

Second, we consider how the result given in Theorem 2 on the approximation of a continuous function by ridge functions can be done in practice. The method is based on the inverse formula for the Radon transform that bears similarity to the inverse formula for Fourier transform. Details and history of this method is discussed e.g. in [54].

Let $f : \mathbb{R}^n \to \mathbb{R}^m$ be a continuous function that is zero outside the ball $B_{\mathbb{R}^n}(0, R) = \{x \in \mathbb{R}^n; |x| < R\}$. Again, $n = 2j_0$.

Let us recall some properties of the Radon transform of a function $f : \mathbb{R}^n \to \mathbb{R}^m$. Let $\omega \in S^{n-1} = \{\omega \in \mathbb{R}^n; \|\omega\| = 1\}$ be the unit vector of $\mathbb{R}^n$ and $s \in \mathbb{R}$. We define the Radon transform of function $f$ to be

$$\mathcal{R}f(\omega, s) = \int_{L(\omega,s)} f(x)\, dV_{n-1}(x) = \int_{x\cdot\omega=s} f(x)\, d^{(n-1)}x \tag{65}$$

where $L(\omega, s) = \{x \in \mathbb{R}^n;\ x \cdot \omega = s\}$ is an $(n-1)$-dimensional hyperplane in $\mathbb{R}^n$ and $dV_{n-1}$ is the $(n-1)$-dimensional volume on this hyperplane.

The function $f$ can be computed from its Radon transform $\widetilde{f}(\omega, s) = \mathcal{R}f$ using the inverse Radon transform that is given by the following formulas (see [53, Section 2, formula (2.5)] or [49])

$$W(\omega, s) = c_n \left(\frac{\partial}{\partial s}\right)^{n-1} \widetilde{f}(\omega, s), \tag{66}$$

$$G(\omega, s) = \mathcal{H}W(\omega, s) = \frac{1}{\pi}\int_{\mathbb{R}} \frac{W(\omega, t)}{s - t}\, dt, \tag{67}$$

$$f(x) = \int_{S^{n-1}} G(\omega, x \cdot \omega)\, d\omega, \tag{68}$$

where $c_n = \frac{1}{2}(2\pi)^{1-n}(-1)^{(n-2)/2}$. Rigorously speaking, above the function $\mathcal{H}W(\omega, s)$, called the Hilbert transform of $W$, is defined by the 1-dimensional Fourier transform $\mathcal{F}$ as $\mathcal{H}w = -i\mathcal{F}^{-1}(\mathrm{sign}\,(\xi) \cdot \mathcal{F}w(\xi))$.

The formula (68) can be used to write an approximation of the function $f$ by ridge functions. Indeed, when function $f(x)$ is given, we first need to compute functions $\widetilde{f}(\omega, s) = \mathcal{R}f$, $W(\omega, s)$ and $G(\omega, s)$ using formulas (65), (66) and (67). Then, let $\omega_1, \omega_2, \ldots, \omega_K \in S^{n-1}$ be $K$ points of the sphere $S^{n-1}$ that form a dense computational grid on $S^{n-1}$. We approximate the measure on the sphere $S^{n-1}$ by point masses with weights $V_k$ at point $\omega_k$, and approximate the integral in (68) by a discrete sum,

$$f(x) = \int_{S^{n-1}} G(\omega, x \cdot \omega)\, d\omega \approx \sum_{k=1}^{K} G(\omega_k, x \cdot \omega_k) V_k. \tag{69}$$

Let now $h > 0$ be small number that determines the distance of computational grid points in $\mathbb{R}$ used below. Next, we approximate the function

$$g_k(s) = G(\omega_k, s), \quad s \in \mathbb{R} \tag{70}$$

by a function that is close to a piecewise constant function having jumps at points $\{mh;\ m \in \mathbb{Z}\} \subset \mathbb{R}$, that is by a weighted sum of smoothed step functions

$$g_k(s) \approx g_k(-Lh) + \sum_{\ell=-L+1}^{L} c_{k,\ell} H_a(s - \ell h) \tag{71}$$

where

$$c_{k,\ell} = g_k(\ell h) - g_k((\ell-1)h), \tag{72}$$

and $H_a(s) = \frac{1}{2}(1 + \text{sign}_a(s))$ is the smoothed Heaviside function and $a > 0$ is small. By doing a simple manipulation of formula (71) we obtain

$$g_k(s) \approx c_k + g_k(-Lh) + \sum_{\ell=-L+1}^{L} \frac{1}{2} c_{k,\ell} \text{sign}_a(s - \ell h), \quad c_k = \sum_{\ell=-L+1}^{L} \frac{1}{2} c_{k,\ell}. \tag{73}$$

Combining formulas (69) and (73) we obtain for $x \in B_{\mathbb{R}^n}(0, R)$

$$f(x) \approx C + \sum_{k=1}^{K} \sum_{\ell=-L+1}^{L} \frac{V_k}{2} c_{k,\ell} \text{sign}_a(x \cdot \omega_k - \ell h), \quad C = \sum_{k=1}^{K} V_k \Big( g_k(-Lh) + c_k \Big).$$

The above Ridge function approximation of $f(x)$ gives an algorithmic way to obtain an approximation corresponding to Theorem 2.

## 2.4  Universal computing by laser based systems

Consider an analogical computing device that takes in $n = 2j_0$ analogical signals, modelled by a point $x = (x_1, x_2, \ldots, x_n)$ in an $n$-dimensional cube $[-R_0, R_0]^n \subset \mathbb{R}^n$, where $R_0 = R/\sqrt{n}$ and $\mathbb{C}^{j_0}$ is identified with $\mathbb{R}^n$, and gives output $f(x)$ that depends continuously on the input $x$. Such a computing device with an output function $f : [-R_0, R_0]^n \to \mathbb{R}^m$ can by Theorem 1 be approximated with an arbitrary precision by a network of lasers having the output function $\mathcal{M}_a(x)$, given by (45). More precisely, and if one sends a signal corresponding to the input $x$ in the network of lasers, waits until the system reaches its steady state, and records the electromagnetic fields coming out from the lasers, one obtains the signal $\mathcal{M}_a(x)$ approximating $F(x)$. This means that the network of lasers, modelled by functions $\mathcal{M}_a : \mathbb{C}^{j_0} \to \mathbb{C}^{j_0}$, are universal approximators. Thus the network of lasers have similar computational capabilities as general neural networks that operate with real valued signals and have sigmoid type activation functions [48].

As any analogical computing device, taking in bounded signals, can be approximated by a network of lasers, also any digital computing device, taking in signals from a finite set of possible inputs, can be approximated by a neural network [45], or in our case, by a network of lasers. Indeed, assume that the possible inputs are a union of disjoint balls $B_{\mathbb{R}^n}(x^{(j)}, \rho) \subset [-R_0, R_0]^n$, $j = 1, 2, \ldots, J$, which centres $S_{input} = \{x^{(1)}, x^{(2)}, \ldots x^{(J)}\}$ and radius $\rho > 0$ are such that $|x^{(j)} - x^{(k)}| > 2\rho$ when $j \neq k$. Here, the input signal $x$ in a ball $B_{\mathbb{R}^n}(x^{(j)}, \rho)$ is interpreted to correspond to the input $x^{(j)}$ with a small error in coding. Also, we assume that the possible outputs are coded by elements $S_{output} = \{y^{(1)}, y^{(2)}, \ldots y^{(J)}\} \subset \mathbb{R}^m$. The digital computing device is modelled by a function $F$ that maps the balls $B_{\mathbb{R}^n}(x^{(j)}, \rho)$ to the points $y^{(j)}$. Then one can construct a network of lasers corresponding to a function $\mathcal{M}_a$ that maps the balls $B_{\mathbb{R}^n}(x^{(j)}, \rho)$ in to the balls $B_{\mathbb{R}^m}(y^{(j)}, \epsilon)$ where $\epsilon > 0$ is arbitrarily small.

In Automata theory, the combinational logic circuits (or the combinational switching circuits) are devices where the output is a function of the present input and inputs are elements of a finite library set, see [50, Section 3.2]. Thus, the above results mean that any combinational logic circuit can be implemented (with arbitrarily small error in the coding of data) by a network of lasers. Due to analysis in Section 1, see formula (32), a laser with a constant time harmonic input approaches a steady state with an exponential speed. This makes it possible to design analog computational gates with finite delays, see [47, Section 11.1], using a network of lasers. As discussed by Minsky in [52], and extended in [59, 62, 67], the finite transition table of any Turing machine [65] can be modelled by a neural network. Hence, if an external memory (or an arbitrary precise storage of binary numbers) is combined with a suitable neural network, the construction of an arbitrary Turing machine becomes possible, see e.g. [60, 61]. Thus the above implies that a network of lasers, combined with an external memory and a clock, can in principle be used to implement an arbitrary Turing machine.

## 3 Simple constructions for OR and AND gates

In this section we document the elementary mathematical model corresponding to our laboratory experiments, and explain in details how a laser can be used as a logical gate.

For an amplitude of an electric field $E$ we use notation

$$E = \begin{pmatrix} E^v \\ E^h \end{pmatrix}$$

where $E^v \in \mathbb{C}$ is the amplitude of the vertical polarization and $E^h \in \mathbb{C}$ is the amplitude of the horizontal polarization. Again, the waves have the carrier waves $\exp(-i(kz - \omega t))$, so that these amplitudes correspond to the time-harmonic electric field $E \exp(-i(kz - \omega t))$.

Let us consider the steady state of one laser with incoming signals

$$E_1 = \begin{pmatrix} (2b_1 - 1) \\ 0 \end{pmatrix}, \quad E_2 = \begin{pmatrix} (2b_2 - 1) \\ 0 \end{pmatrix}, \tag{74}$$

where

$$b_1, b_2 \in \{0, 1\} \tag{75}$$

corresponds to bits "zero" or "one". Then the amplitudes $a_j = (2b_j - 1)$ satisfy $a_j \in \{-1, 1\}$. We also use a bias signal

$$E_0 = \begin{pmatrix} a_0 \\ 0 \end{pmatrix}$$

where $a_0 \in \{-1, 1\}$. Let us assume that the signals are phase-controlled and we can mix them so that the light going in a laser has the amplitude

$$E_{in} = E_1 + E_2 + E_0.$$

Then output amplitude from the laser is

$$E_{out}(t) = \frac{E_{in}(t)}{|E_{in}(t)|} = \begin{pmatrix} a_{out} \\ 0 \end{pmatrix},$$

and we define the logical variable corresponding to the output amplitude to be

$$b_{out} = \frac{1}{2}(a_{out} + 1).$$

First, consider the case when the bias signal has the amplitude $a_0 = -1$. Then $b_{out} = 1$ if $b_1 = b_2 = 1$ and in other cases $b_{out} = 0$. Thus the laser, operating in the steady state, can be considered as an AND gate.

Second, consider the case when the bias signal has the amplitude $a_0 = 1$. Then $b_{out} = 0$ if $b_0 = b_2 = 0$ and in other cases $b_{out} = 1$. Thus the laser, operating in the steady state, can be considered as an OR gate.

This simple reasoning shows that in the steady state the laser works with inputs (74)-(75) as the AND gate if the bias amplitude is $-1$ and as the OR gate if the bias amplitude is $+1$. Hence a laser can be considered as a programmable logical gate. As the NOT gate corresponds to a phase shift that can be implemented by choosing the length of an optical path that connects two logical units appropriately, we see that any Boolean function can be implemented using a multi-level network of lasers. We note that such multi-level network the lasers, where information propagates from layer number $n$ to the number $(n+1)$ and so on, the optical pathways connecting different layers need to have optical isolators, or the power of lasers needs to decrease from each layer to the next one.

Also, by connecting lasers in various ways we can also design systems where the coding of the signals are more flexible, e.g., based on amplitude or polarisation coding and also logical gates where the signals are not phase controlled.